\documentclass[preprint,11pt,authoryear]{elsarticle}  
\usepackage{amsmath, amssymb, subfigure, bm}
\usepackage{graphicx}
\usepackage{supertabular}
\usepackage{longtable}

\journal{Icarus}

\begin{document}

\begin{frontmatter}  


\title{Two Dynamical Classes of Centaurs}


\author[pam]{Brenae L. Bailey}
\ead{bbailey@math.arizona.edu}

\author[lpl]{Renu Malhotra}
\ead{renu@lpl.arizona.edu}

\address[pam]{Program in Applied Mathematics, 617 N.~Santa Rita, The University of Arizona, Tucson, AZ 85721}
\address[lpl]{Lunar and Planetary Laboratory, 1629 E.~University Blvd., The University of Arizona, Tucson, AZ 85721
}

\begin{abstract}
The Centaurs are a transient population of small bodies in the outer solar system whose orbits are strongly chaotic.   These objects typically suffer significant changes of orbital parameters on timescales of a few thousand years, and their orbital evolution exhibits two types of behaviors described qualitatively as random-walk and resonance-sticking.  We have analyzed the chaotic behavior of the known Centaurs.  Our analysis has revealed that the two types of chaotic evolution are quantitatively distinguishable: (1) the random walk-type behavior is well described by so-called generalized diffusion in which the rms deviation of the semimajor axis grows with time $t$ as $\sim t^H$, with {\it Hurst} exponent $H$ in the range 0.22--0.95, however (2) orbital evolution dominated by intermittent resonance sticking, with sudden jumps from one mean motion resonance to another, has poorly defined $H$.  We further find that these two types of behavior are correlated with Centaur dynamical lifetime: most Centaurs whose dynamical lifetime is less than $\sim$ 22 Myr exhibit generalized diffusion, whereas most Centaurs of longer dynamical lifetimes exhibit intermittent resonance sticking.  We also find that Centaurs in the diffusing class are likely to evolve into Jupiter-family comets during their dynamical lifetimes, while those in the resonance-hopping class do not.

\end{abstract}





\end{frontmatter}

\section{Introduction}
\label{intro}
The Centaurs are a dynamical class of small bodies in the outer solar system whose orbital parameters lie in a range intermediate between those of the Kuiper belt and of the Jupiter-family comets.  Numerical simulations have found that the Centaurs are typically removed from the solar system on timescales of only a few million years \citep{LD1997,DLD1996, TM2003, HEB2004a,DB2007}
. These dynamical lifetimes are very short compared to the age of the solar system, implying that the Centaurs are a transitional population with a source elsewhere in the system. Likely source populations are the several dynamical subclasses of the Kuiper belt beyond Neptune \citep{LD1997,VM2008}.  Possible sinks of the Centaur population include the Kuiper belt's scattered disk, the Jupiter-family comets, and the Oort cloud; Centaurs are also removed from the solar system by ejection on hyperbolic orbits or by collisions with a planet. Numerical analysis of their orbital evolution shows that these objects typically suffer frequent close encounters with the giant planets and their orbits are strongly chaotic. 

Previous studies have provided detailed qualitative descriptions of the different types of chaotic behavior of Centaurs. The present study aims for a quantitative analysis of Centaur chaotic dynamics by using a generalized diffusion approach, which is a relatively new tool in solar system dynamics.  Towards this end, we started with the known sample of Centaurs, and we carried out a 100 million year (Myr) numerical integration of their orbits under the perturbing influence of the four giant planets. (The length of this integration is more than ten times the median dynamical lifetimes of the observed Centaurs found in previous studies.) We then analyzed the Centaurs' fluctuations in semimajor axis to determine how the root mean square fluctuations evolve over time.  We found two distinct types of evolution of the semimajor axis fluctuations.  The first is characterized by diffusion-like evolution, wherein the mean square fluctuations of the semimajor axis increase as a power law of time. The second type does not show this power law behavior; instead, the fluctuations increase slowly at short timescales and more rapidly at larger timescales, suggesting that multiple processes are at work.  These two types of behavior are strongly correlated with Centaur lifetime: with few exceptions, Centaurs exhibiting the diffusion-like behavior have dynamical lifetimes shorter than $\sim22$ Myr, whereas the second type have longer dynamical lifetimes.  We also find that the latter group of Centaurs are strongly correlated with the `resonance-sticking' behavior noted qualitatively in previous studies, in which Centaurs become temporarily trapped in mean motion resonances with the giant planets; these Centaurs typically hop from one resonance to another for much of their dynamical lifetimes. Our analysis shows that the two types of behavior can be objectively and quantitatively distinguished, and suggests that the Centaurs may be comprised of two distinct dynamical classes.  

This paper is organized as follows.  We summarize previous studies of Centaur dynamics in section \ref{previous}.  In section \ref{sim}, we describe the numerical simulations we have carried out to explore the orbital evolution of Centaurs.  Section \ref{anal} describes the analyses that we have applied to the results of our simulations. In section \ref{conc} we provide a summary and conclusions.

\section{Previous Work}
\label{previous}

 The Minor Planet Center (MPC) defines Centaurs  as ``objects [with] perihelia beyond the orbit of Jupiter and semimajor axes inside the orbit of Neptune''\footnote{http://www.cfa.harvard.edu/iau/lists/Unusual.html}.
The MPC provides a combined list of Centaurs and scattered disk objects. (The latter are not defined on the MPC's webpages but are generally identified as objects with perihelia near or slightly beyond Neptune's orbit and semimajor axes greater than 50 AU.) While there is a general consensus on the lower bound for Centaur orbits (although \citet{GMV2008} propose a higher cutoff of $q > 7.35$ AU to exclude objects whose dynamics are controlled by Jupiter), the upper bound is open for interpretation.  The samples of Centaurs studied by different authors thus vary slightly based on the authors' chosen criterion, with some authors constraining semimajor axis only and others using constraints based also on perihelion and aphelion distance. 

The first Centaur, Chiron, was discovered in 1977 \citep{K1989}, but only in the past decade or so has there been sufficient computing power to explore the dynamical behavior of Centaurs using large-scale integrations of particle orbits.   \citet{DLD1996} simulated the orbital evolution of $\sim800$ particles ``cloned'' from the six Centaurs known at that time that had values of semimajor axis $a$ between 6 and 25 AU. The initial orbital elements of each particle were identical to those of one of the Centaurs, save for the addition of a random variation in $a$ of order $10^{-5}$ AU.  They focused on the dynamical lifetimes of the particles, and found median lifetimes of 0.5 to 5 Myr for their six ensembles of clones.  Lifetime was most sensitive to perihelion distance: at smaller perihelion, particles are more likely to encounter one of the more massive planets and receive large gravitational ``kicks'', and thereby be removed from the Centaur population.  They also reported that the number of surviving particles in each ensemble decreased at first roughly exponentially with time, then more slowly as a power law.

In work published in 1997, Levison and Duncan \nocite{LD1997} followed the evolution of a sample of hypothetical Kuiper belt objects as they evolved their way inward toward the Sun to become Jupiter-family comets.  A subset of their sample thus spent time as Centaurs.  These authors argued that a dynamical classification based on Tisserand parameter $T_p$ is more appropriate for objects on planet-encountering orbits than the traditional divisions based solely on semimajor axis (or equivalently, orbital period).  The Tisserand parameter is defined in the context of the circular restricted 3--body problem, and is given by
\begin{equation}
	T_p = \frac{a_p}{a}+2\sqrt{\frac{a}{a_p}(1-e^2)}\cos{i},
\label{eq:tisserand}
\end{equation}
where $a_p$ is the semimajor axis of the planet, $a$ and $e$ are the semimajor axis and eccentricity of the small body in the heliocentric frame, and $i$ is the inclination of the small body relative to the orbit of the planet.  The Tisserand parameter is nearly constant for a given particle before and after an encounter with a planet. Note that when $i=0$, $e=0$, and $a=a_p$, then $T_p = 3$, so values of $T_p$ near 3 indicate that the orbit of the particle is similar to the orbit of that planet (although it is not guaranteed), and the planet can strongly influence the orbit of the particle.  In particular, Levison and Duncan identify the Centaurs with what they call Chiron-type comets, defined by $T_J>3$ and $a>a_J$, where $T_J$ is the Tisserand parameter with respect to Jupiter and $a_J=5.2$ AU is the semimajor axis of Jupiter.  Jupiter-family comets are defined as objects with $2<T_J<3$.
	
Levison and Duncan's results suggested that Centaurs can become Jupiter-family comets by being ``handed'' inward from one planet to the next through a series of close encounters.  Based on the approximate conservation of $T_p$ and their assumed initial values of $a$, $e$, and $i$ for their hypothetical source population in the Kuiper belt, they calculated that, starting with Neptune, each planet could scatter a small body just far enough inward to reduce its perihelion distance so that the body could cross the orbit of the next planet in.  About 30\% of the particles in their integrations did become Jupiter-family comets. 

\citet{TM2003} carried out a study of the dynamics of all of the known Centaurs as of 2002.  Their numerical simulation included the four giant planets and 53 Centaurs, the latter treated as massless test particles, and they followed the orbits for 100 Myr. They chose their sample of Centaurs based on perihelion distance $q$ alone, using the criterion $5.2 < q < 30$ AU. Particles were removed from the simulation when they reached either $r > 20,000$ AU or $r < 2.5$ AU where $r$ is the heliocentric distance.  From this simulation, they found a median dynamical lifetime of 9 Myr, longer than but on the same order as the lifetimes found by \citet{DLD1996}.  Only 7 of the 53 particles (13\%) survived the full 100 Myr integration.  


Tiscareno and Malhotra also observed qualitatively different types of behavior over time, as indicated by time series plots of semimajor axis for different particles.  They described these as ``resonance hopping'' and ``random fluctuations,'' with some particles exhibiting a combination of both.  This work provides the motivation and starting point for our own work, described in section \ref{sim}. 

 \citet{HEB2004a} have also investigated the dynamical evolution of Centaurs. They integrated the orbits of 23328 particles cloned from 32 Centaurs for 3 Myr both forward and backward in time.  From their results, they extrapolate half-lives of 0.5 to 32 Myr for each Centaur, defined as the time when half of the ensemble of clones of that Centaur have been removed from the simulation by either reaching a heliocentric distance of 1000 AU or colliding with a massive body. They also estimate a total population  of $\sim44300$ Centaurs with diameters greater than 1 km, based on the fraction of particles in their simulation that become short-period comets and an assumed flux of one new short-period comet every 200 years. For this calculation they consider Centaurs to be objects with perihelion $q > 4$ AU and aphelion $Q < 60$ AU.  The authors propose a classification scheme for Centaurs based on both perihelion and aphelion, suggesting that whichever planet is nearest at those parts of the orbit controls the dynamics of the Centaur.  This results in 18 dynamical classes for objects with perihelia between 4 and 33.5 AU.  While the defining assumption is reasonable for a very detailed description, the number of categories is probably too large for the purpose of describing the big picture of Centaur dynamics. 

\citet{DB2007} modeled the origin of the Centaurs from a presumed source in the trans-Neptunian scattered disk population.  They generated initial conditions of the source population by debiasing the orbital element distribution of 95 observed scattered disk objects, and then performed a numerical integration of 1000 test particles for 100 million years. From this simulation, they tracked particles that evolved into the Centaur zone and they estimated the intrinsic population and the orbital distribution of Centaurs.  They gave a comprehensive description of the qualitatively different types of evolution found in their simulation. Furthermore, they determined a mean lifetime of 72 Myr for Centaurs, and a very strong and smooth dependence of lifetime on perihelion distance. The authors explained qualitatively that their large value of the mean lifetime (compared with previous studies) was owed to a larger fraction of their sample exhibiting either resonance hopping or pseudo-stable states with near-conservation of perihelion distance in the range between Saturn's and Neptune's orbits.

In the present work, we do not attempt to trace the origins of Centaurs or to determine the distribution of their dynamical lifetimes.  Rather, our goal is to make a step towards a quantitative analysis of the qualitative descriptions of Centaurs' chaotic orbital evolution reported in previous studies.  We make use of the tools of generalized diffusion (section \ref{diff}), which is a relatively  new application in solar system dynamics.  Although we analyzed only a small sample (63 observed Centaurs), our results show that the qualitative behaviors found in previous studies are objectively quantifiable by means of a Hurst exponent.

\section{Simulations}
\label{sim}
Our orbital integrations were done using the RA15 integrator \citep{E1985}, a 15th-order variable step size method for ordinary differential equations, which is part of the public-domain software package \textit{Mercury} \citep{C1999}, designed for N-body integrations in planetary dynamics applications.  Everhart's orbit integrator has often been used for cometary orbits and is very stable for large eccentricities and close planetary encounters; the price for its high accuracy is its larger computational time requirement, compared to the lower accuracy hybrid symplectic method also offered in the \textit{Mercury} package.  As numerical accuracy was a priority for the analysis of the strongly chaotic orbit evolution, and our simulation involved only a small number of known Centaurs, the penalty in computational time was not prohibitive.  We used a relative position and relative velocity error tolerance of $10^{-12}$.

Our primary simulation included the Sun, the four giant planets, and 63 Centaurs treated as massless test particles.  The sizes, masses, and initial positions of the Sun and the giant planets were obtained from JPL Horizons service\footnote{http://ssd.jpl.nasa.gov/?horizons}.  
The initial conditions for the Centaurs were taken from the Minor Planet Center's online list of Centaurs and Scattered Disk Objects\footnote{http://www.cfa.harvard.edu/iau/lists/Centaurs.html} on 6 March 2007.  Our sample is a subset of that list, selected based on the criterion $q>5.2$ AU and $a<30$ AU. Note that this is slightly different from the criterion used by  \citet{TM2003} and \citet{DB2007}, and excludes more fully the scattered disk objects. Table \ref{table:ics} lists the Centaurs in our sample with their initial conditions and length of observations.  This sample includes some objects with short observational arcs from single oppositions, so the given orbital elements for those objects have higher uncertainties.

\marginpar{\ \ \ \bf Fig.~1}

\begin{figure}[tbh]
	\centering
		\includegraphics[width=1.00\textwidth]{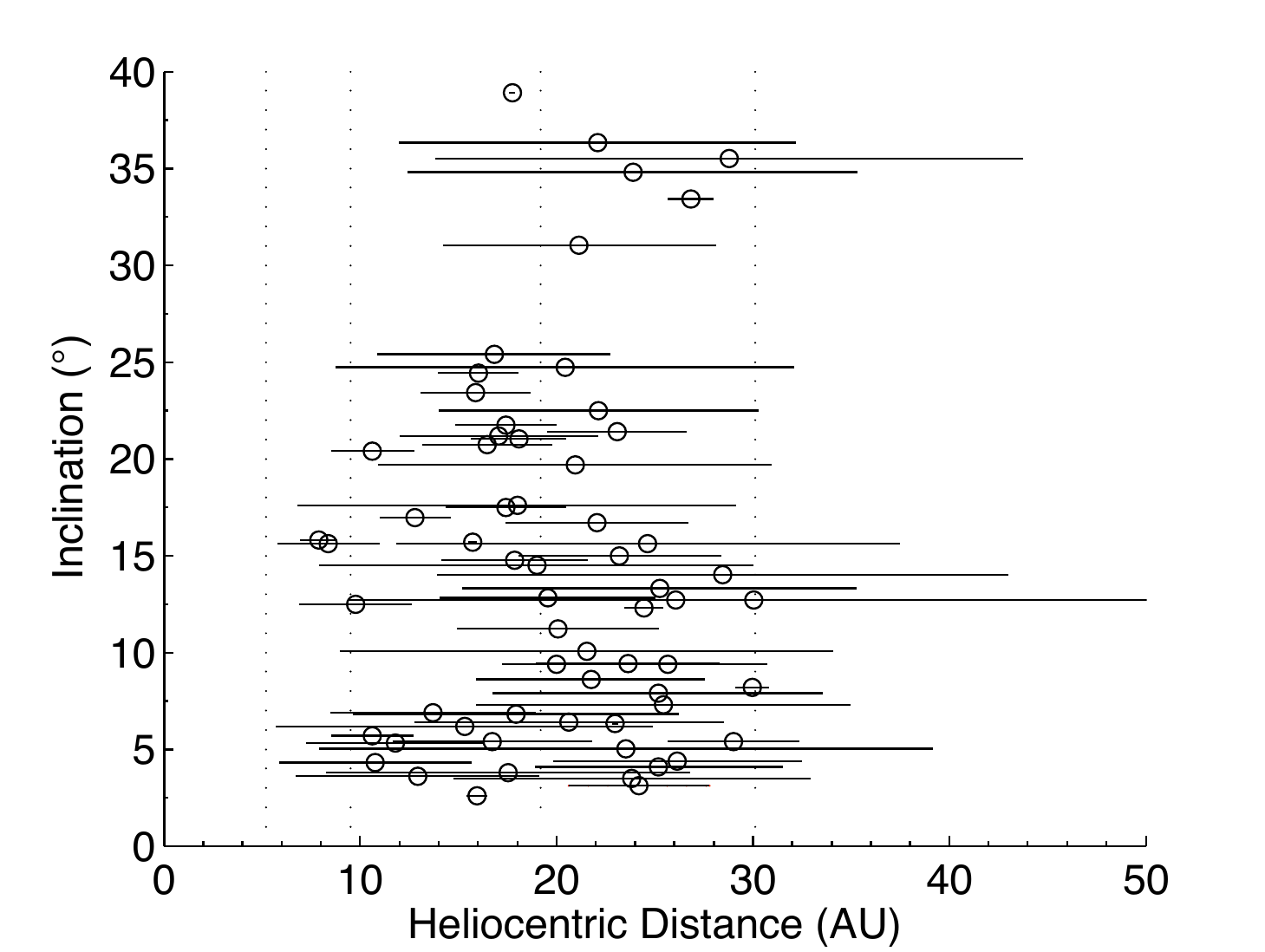}
		\caption{The initial orbital elements for our sample of Centaurs. The circles indicate the semimajor axes and the horizontal bars extend from perihelion to aphelion.  }
	\label{fig:initconds}
\end{figure}

A plot of the initial semimajor axes, eccentricities and inclinations of our sample of Centaurs is shown in Fig.~\ref{fig:initconds}.  The circles mark the semimajor axes and the horizontal bars on the plot extend from perihelion to aphelion for each Centaur, illustrating the eccentricity of each orbit.  
As seen in the figure, the initial inclinations of the Centaurs span the range $3^{\circ}<i<40^{\circ}$. Their eccentricities range from $\sim0.01$ to 0.68.

 We integrated the orbits of the planets and Centaurs for 100 Myr and recorded the evolved orbital elements every 300 years.  Centaurs were removed from the simulation when they either reached a heliocentric distance of $10^4$ AU or collided with a massive object (planet or Sun); we refer to the former as `ejected'.

\marginpar{\ \ \ \bf Fig.~2}

Our simulations confirmed the two types of behavior noted by \citet{TM2003} and \citet{DB2007}.  Sample results are shown in Fig.~\ref{fig:twosamples}.   In the top panel, 2002 CB249 (initial $a = 28.45$ AU, $e = 0.511$, $i = 14.0^{\circ}$) follows a random walk in semimajor axis. In the bottom panel, the orbital evolution of 1998 TF35 (initial $a = 26.09$ AU, $e = 0.378$, $i = 12.7^{\circ}$) is dominated by resonance hopping.  Note that these two objects have quite similar orbits but very different long term dynamics.  We emphasize that due to the chaotic nature of the orbital evolution for all Centaurs, these plots should not be taken as predictions of the actual future evolution of particular objects, but only as examples of the types of behavior that can occur.  Both objects reach semimajor axes $a>30$ AU during their lifetimes, thus leaving the Centaur region. 

Of our initial sample of 63 test particles, all but one spent part of their lifetimes as members of other dynamical classes, including scattered disk objects, resonant Kuiper belt objects, and Jupiter-family comets.  The exception (2006 RJ103) was identified as a Neptune Trojan and we discarded it from further analysis. Nine others survived the full 100 Myr integration.  A histogram of the dynamical lifetimes of our sample of particles is shown in Fig.~\ref{fig:lifetimes}.
The first large gap in lifetimes occurs between 22 Myr and 38 Myr.  Based on this gap, we have designated all particles that survived more than 22 Myr as ``long-lived'' particles, a total of 15 objects. One of these (2005 TH173) appeared to be in a quasi-stable orbit between Saturn and Uranus for almost the entire integration; we describe this exceptional case in section \ref{exceptions}.  Those particles that survived less than 22 Myr before being removed from the simulation are designated as ``short-lived''.

\marginpar{\ \ \ \bf Fig.~3}

The median lifetime for particles in our simulation was 6 Myr, similar to the value of 9 Myr found by  \citet{TM2003} and close to the longest dynamical lifetimes found by \citet{DLD1996}.  The larger value found by Tiscareno and Malhotra is likely due to the fact that they included objects with initial $a>30$ AU, which likely belong in the scattered disk classification and may have survived longer due to fewer encounters with planets.  
Similarly, the Centaurs studied by \citet{DLD1996} had smaller perihelia and therefore shorter lifetimes, a correlation first reported in \citet{DB2007}. 

\section{Analysis}
\label{anal}
In this section we describe our analysis of the two distinct types of behavior seen in the results of our simulation: random walks and resonance hopping. 

\subsection{Generalized Diffusion}
\label{diff}
If a particle undergoes a random walk with either fixed or normally distributed stationary independent increments, its mean square displacement from the origin $\left\langle x^2\right\rangle$ at time $t$ grows linearly with $t$. In the limit as the step sizes approach 0, this process leads to Brownian motion \citep{E1905}. Recent work has extended this framework to {\it generalized diffusion} \citep{WS1998,MK2000,C2006}, in which 
\begin{equation}
\left\langle x^2\right\rangle = Dt^{2H}, \quad \hbox{for } 0<H<1,
\end{equation}
where $D$ is a generalized diffusion coefficient and $H$ is called the Hurst exponent.  In the case of $H\neq \frac{1}{2}$, the random process is called {\it anomalous diffusion}, and it occurs if the steps are correlated in some way. The degree of correlation is related to the deviation of $H$ from the classical diffusion value of $\frac{1}{2}$.

 As noted above, many of the particles in our simulations appear to follow a random walk in semimajor axis.  We have analyzed the diffusion characteristics of our sample of Centaurs by calculating the Hurst exponent of each Centaur based on the time series of its semimajor axis. Since the orbital energy per unit mass is related to semimajor axis ($E \propto \frac{1}{a}$), $a$ is a proxy for the orbital energy and provides a useful measure of the orbital evolution.  The steps in our analysis for orbital diffusion of Centaurs are as follows:
 \begin{enumerate}
	\item Choose a window length $w$, corresponding to a fixed time interval.  
	\item Apply overlapping windows of length $w$ to the data set. Each window was allowed to overlap its neighbors by half its length. If the amount of data not included in any window was greater than $\frac{1}{4}$ the window length, an additional window was applied to cover the end of the data set. 
	\item Calculate the standard deviation $\sigma$ of $a$ within each window.
	\item Find the average standard deviation $\bar\sigma(w)$ for all windows of length $w$.
	\item Repeat steps 1-4 for different window lengths.
	\item Plot log~$\bar\sigma(w)$ vs.~log~$w$.  The slope of the best-fit line is an estimate of the Hurst exponent, $H$. 
\end{enumerate}
The window lengths were chosen by dividing each data set uniformly into 16 logarithmic bins, from 3000 years up to the length of the data set. The full length of the data set was used as the upper bound in order to minimize the loss of coverage due to rounding.  Any windows larger than $25\%$ of the data set were then discarded to minimize errors in calculating $\bar\sigma(w)$. 

For example, an object that survived 3 Myr would have $(3 \times 10^6 / 300) + 1 = 10001$ points in the data set, since data were recorded every 300 years beginning with time 0.  The minimum window length is $3000/300 = 10$ data points. To find the window lengths for this object, the range $[log_{10} 10, log_{10} 10001]$ would be divided into 16 equal increments, which would then be converted back to numbers of data points and rounded to the nearest whole number.  In years, this gives the set of window lengths $\{3000,$ 4800, 7500, 12000, 18900, 30000, 47400, 75300, 119400, 189300, 300000, 475500, 753600, 1194300, 1893000, $3000300\}$.  The last 4 values are greater than $3 \times 10^6 / 4 = 750000$ and would be discarded.  

This procedure produced between 8 and 14 window lengths for objects in our sample of observed Centaurs. 


\marginpar{\ \ \ \bf Fig.~4}

\begin{figure}[p]
	\centering
	\includegraphics[width=\textwidth]{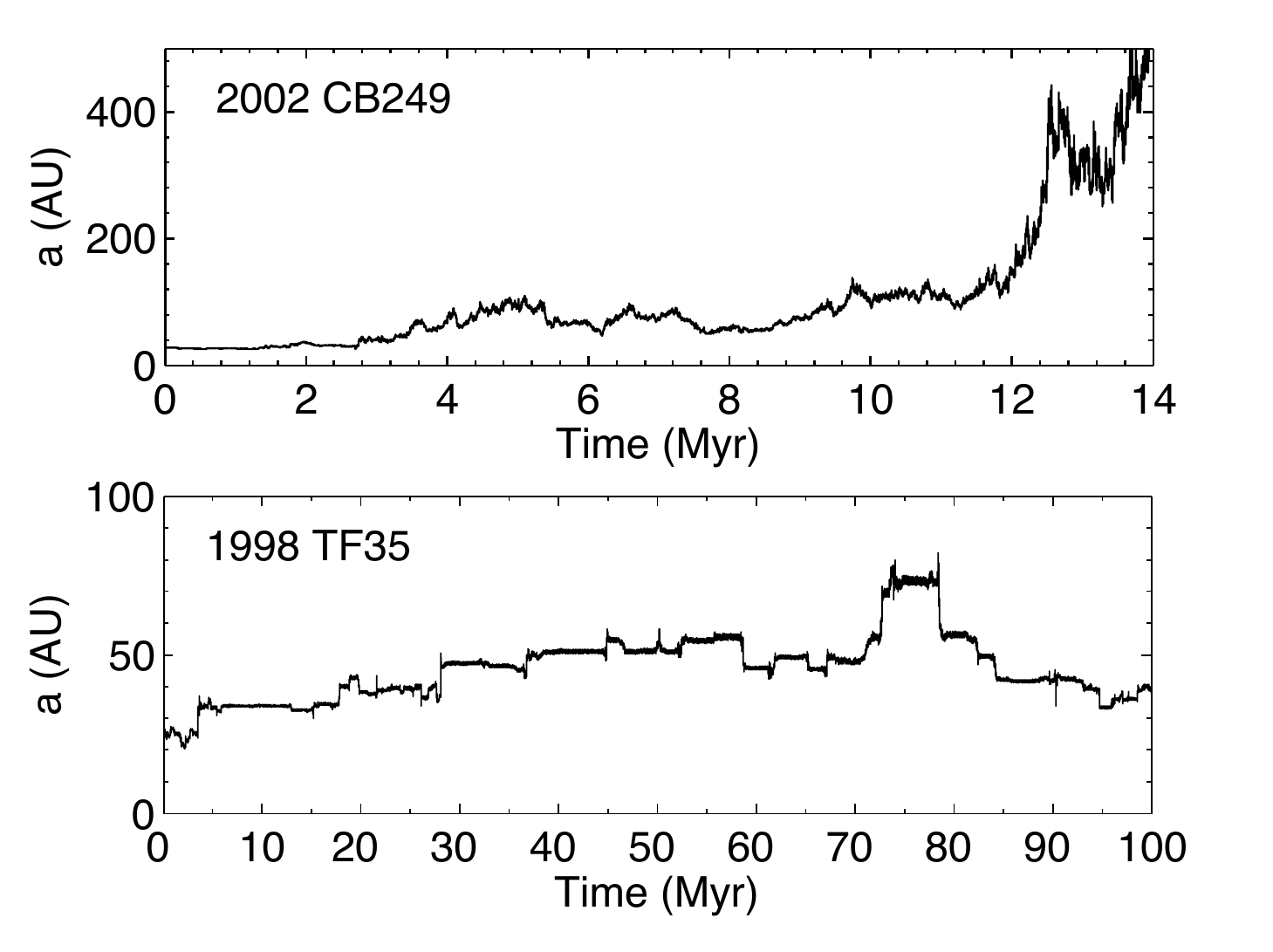}
	\caption{Two examples of Centaur orbital evolution.  Top panel: a particle undergoing a random walk; bottom panel: a particle engaged in resonance hopping.  Note the different scales for the two panels. }
	\label{fig:twosamples}
\end{figure}

\begin{figure}[htb]
	\centering
		\includegraphics[width=1.00\textwidth]{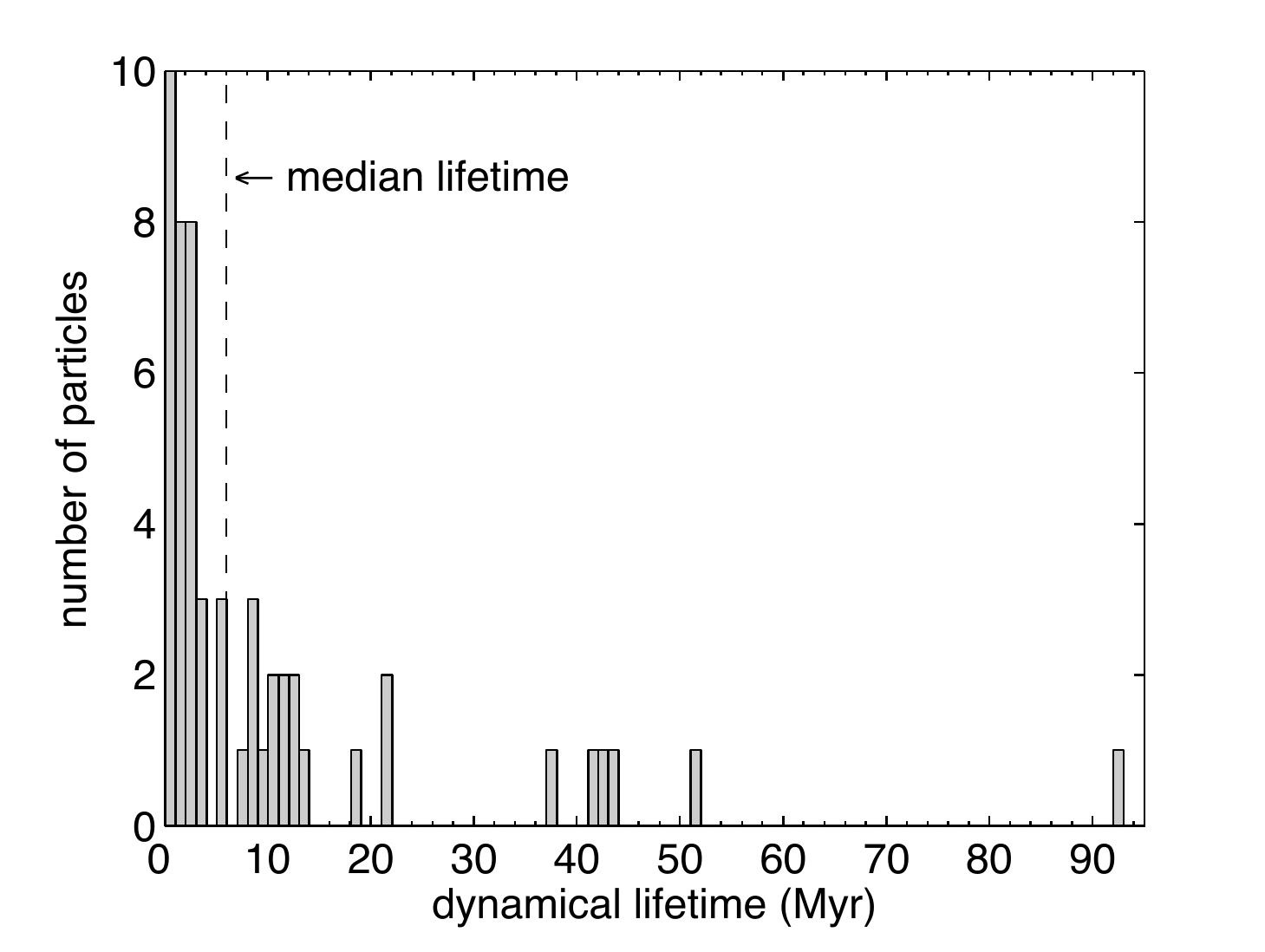}
	\caption{The distribution of dynamical lifetimes for our sample of particles.  The nine Centaurs that survived the full 100 Myr integration are not shown.}
	\label{fig:lifetimes}
\end{figure}

\begin{figure}[p]
	\centering
		\includegraphics[width=\textwidth]{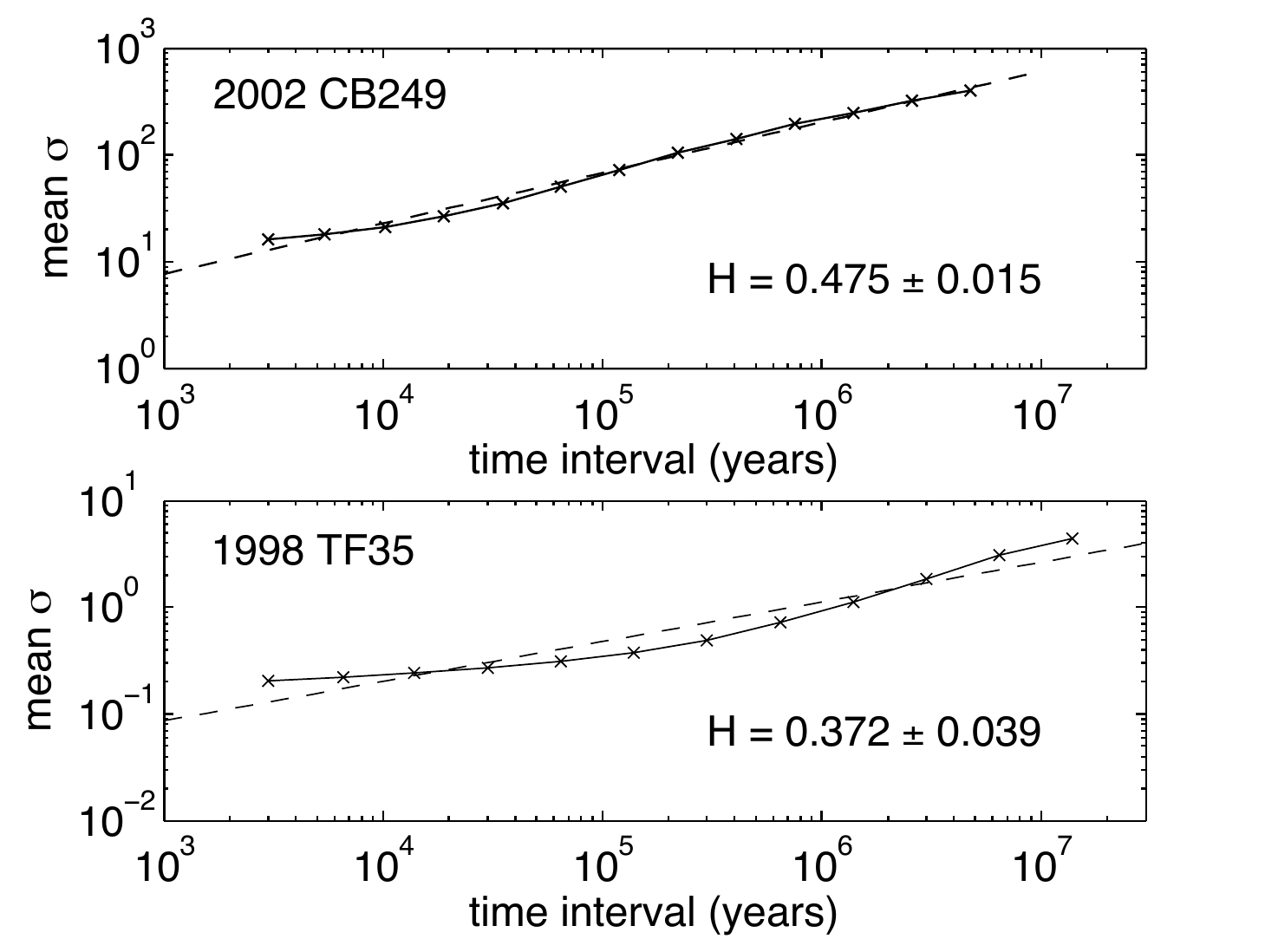}
	\caption{Log-log plots of $\bar\sigma$ vs.~$w$ for the two sample Centaurs shown in Fig.~\ref{fig:twosamples}.  The dotted lines indicate the best-fit linear function. The uncertainties in $H$ are quoted with 1 standard deviation.}
	\label{fig:twoanalyses}
\end{figure}

Results of this calculation for the two Centaurs shown in Fig.~\ref{fig:twosamples} are presented in Fig.~\ref{fig:twoanalyses}.  The upper panel illustrates an example of generalized diffusion, in which the plot of log $\bar\sigma$ vs.~log $w$ is well fitted by a straight line; in this case, the calculated Hurst exponent is $H = 0.48$.  In the lower panel, the dependence of log $\bar\sigma$ on log $w$ is smoothly curved rather than linear, so $H$ is poorly defined in this case.  

\marginpar{\ \ \ \bf Fig.~5}

\begin{figure}[p]
	\centering
		\includegraphics[width=\textwidth]{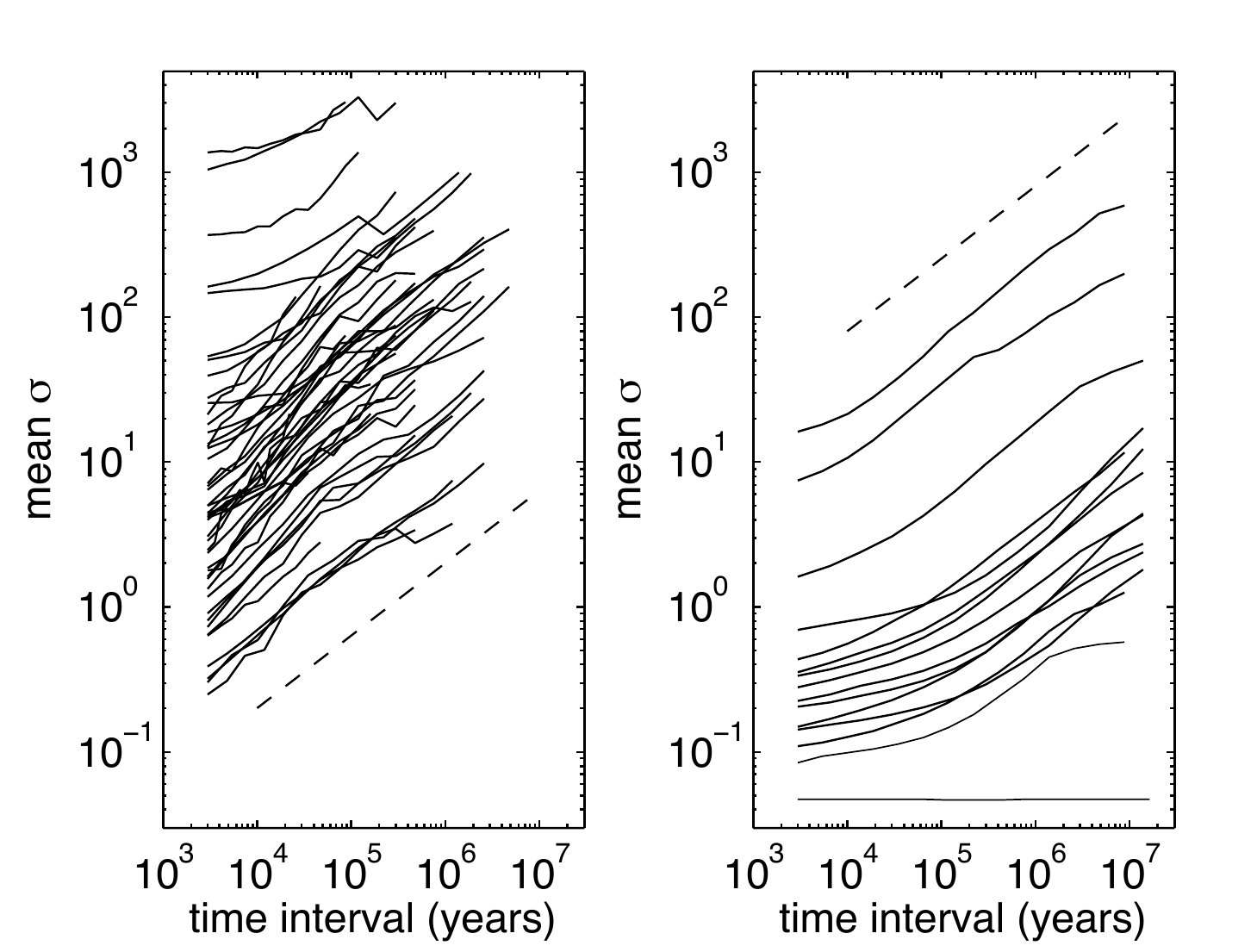}
	\caption{Log-log plots of $\bar\sigma$ vs. $w$ for particles in our simulation. Left panel: short-lived Centaurs; right panel: long-lived Centaurs. The dotted line in each panel is a reference line with slope $H = \frac{1}{2}$. }
	\label{fig:slopes}
\end{figure}

Our results for this calculation for all the Centaurs in our simulation are shown in Fig.~\ref{fig:slopes}. In the left-hand panel, the plots of $\log\bar\sigma$ vs.~$\log{w}$ for the short-lived Centaurs are generally well fitted by lines of constant slope, indicating that generalized diffusion is a good model for the orbital evolution of these particles. We quantify the goodness of fit by calculating the best-fit linear function to the $\log\bar\sigma$ vs.~$\log{w}$ data using a least-squares regression and examining the residuals, a measure of the deviations from the best-fit line.  The linear function is a good fit if the residuals satisfy $|R|<0.08$.  The values of $H$ for this group range from 0.22 to 0.95, with mean 0.56 and standard deviation 0.15. 

In the right-hand panel of Fig.~\ref{fig:slopes}, the plots of $\log\bar\sigma$ vs.~$\log{w}$ for most of the long-lived Centaurs are smoothly curved rather than straight.  Note also that the values of $\bar\sigma$ are typically lower for this group of Centaurs than for the group in the left panel.  
 The ``curved'' group have residuals to a best-fit linear function that exceed $|R| = 0.08$ with a clear pattern of positive residuals for the lowest and highest values of $\log{w}$ and negative residuals for central values, indicating that the $\log\bar\sigma$ vs.~$\log{w}$ curve lies above the best-fit line at the ends and below it in the center, as illustrated in the lower panel of Fig.~\ref{fig:twoanalyses}. These Centaurs' evolution is not well described by a generalized diffusion process. We discuss this further in section \ref{hopping}.  
 
 Two objects in the right-hand panel of Fig.~\ref{fig:slopes} lie well above the rest, indicating large values of $\bar\sigma$ for all time intervals.  These objects are discussed in section \ref{exceptions}.  In all, 51 objects from our sample have well-defined Hurst exponents from the best-fit lines to their $\log\bar\sigma$ vs.~$\log{w}$ plots.  We call this group the diffusing class.  The median lifetime for this class is 3.1 Myr, but this class includes some members with dynamical lifetimes greater than 100 Myr.


\marginpar{\ \ \ \bf Fig.~6}

\begin{figure}[p]
	\centering
		\includegraphics[width=0.9\textwidth]{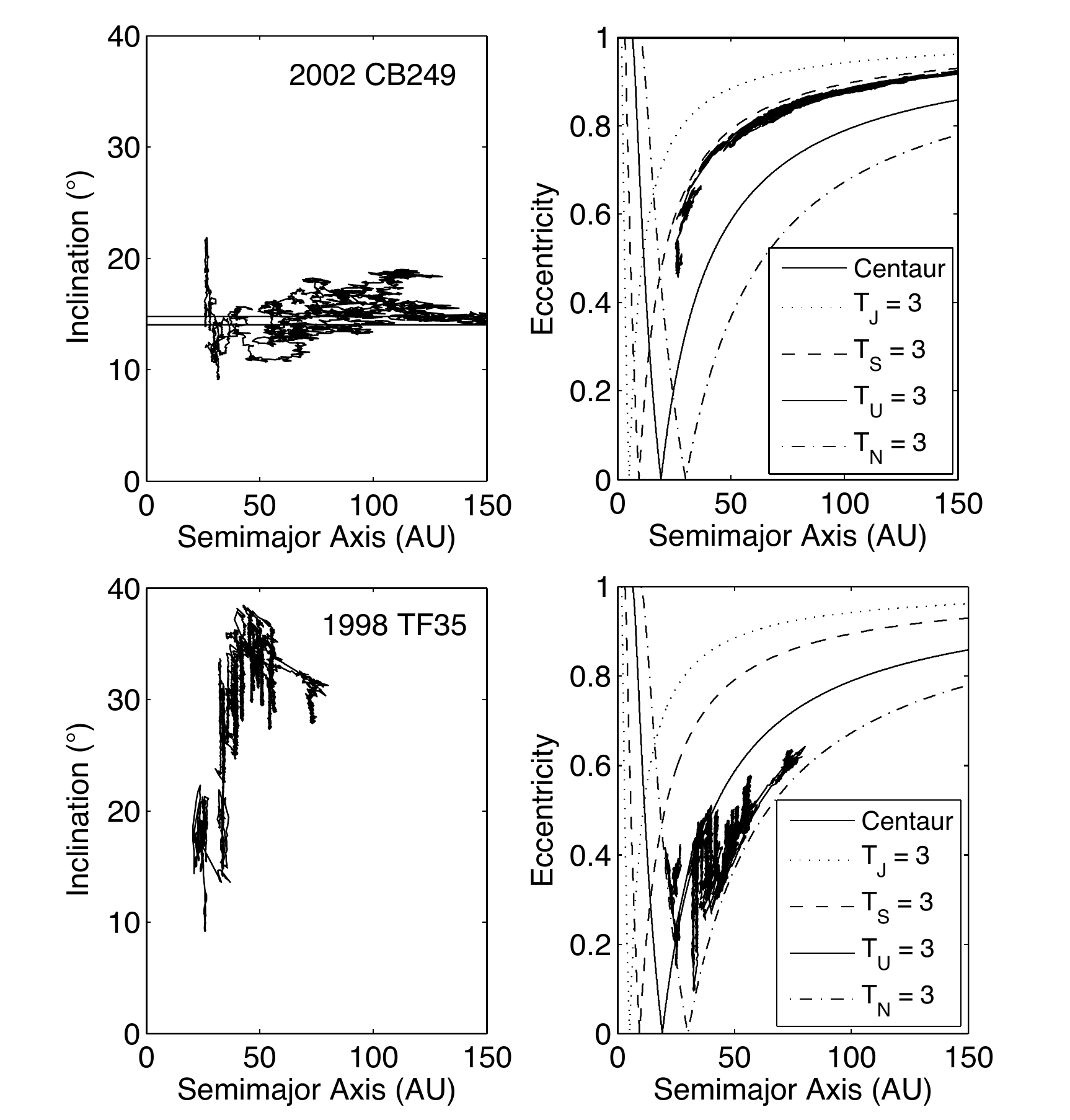}
		\caption{Plots of $i$ vs. $a$ and $e$ vs. $a$ for two Centaurs. Top: 2002 CB249, a short-lived, diffusion-dominated Centaur; Bottom: 1998 TF35, a long-lived, resonance-hopping Centaur. In the right-hand panels, the overlying curves indicate where the Tisserand parameter $T=3$ with respect to Jupiter, Saturn, Uranus, or Neptune, for zero inclination orbits.}
	\label{fig:ea}
\end{figure}

One mechanism for the diffusion in $a$ is suggested by the plots in the upper panels of Fig.~\ref{fig:ea}, which shows the traces in $(a,e)$ and $(a,i)$ planes.  This object, 2002 CB249, is also shown in the top panel of Fig.~\ref{fig:twosamples}.  The plot of $e$ vs.~$a$ shows that this particle spends essentially all of its dynamical lifetime at constant perihelion (near Saturn's orbit), being pumped to higher and higher values of eccentricity until it is ejected from the solar system. 

In contrast, as shown in the bottom panels, 1998~TF35 wanders through a small region of the $(a,e)$-plane, but never exceeds $e \approx 0.5$.  We also see a contrast in the inclination evolution: 1998~TF35 visits a much wider range of inclinations compared with 2002~CB249. The vertical features indicate times spent in resonance.  This pattern is characteristic of resonance hopping, discussed below.  

\subsection{Resonance Hopping}
\label{hopping}

Of the 15 long-lived Centaurs in our sample, 10 exhibit nonlinear curves of $\log\bar\sigma$ vs.~$\log{w}$, as seen in Fig.~\ref{fig:slopes}.  At small values of $w$ (i.e., small timescales), the asymptotic slopes of these curves approach 0.06-0.27, with a mean of 0.15.  These low values reflect the small variations in semimajor axis on timescales of up to $\sim 10^4$ years.  At the high range of values of $w$ (i.e., timescales of $10^6$ years or more), the asymptotic slopes are higher, 0.23-0.85, with mean 0.57; these values are similar to the range of the Hurst exponent for the short-lived Centaurs.  This suggests that the generalized diffusion model may still be applicable for this long-lived group, but only over much longer timescales than for the short-lived group. 

\marginpar{\ \ \ \bf Fig.~7}

\begin{figure}[p]
	\centering
		\subfigure{\includegraphics[width=\textwidth]{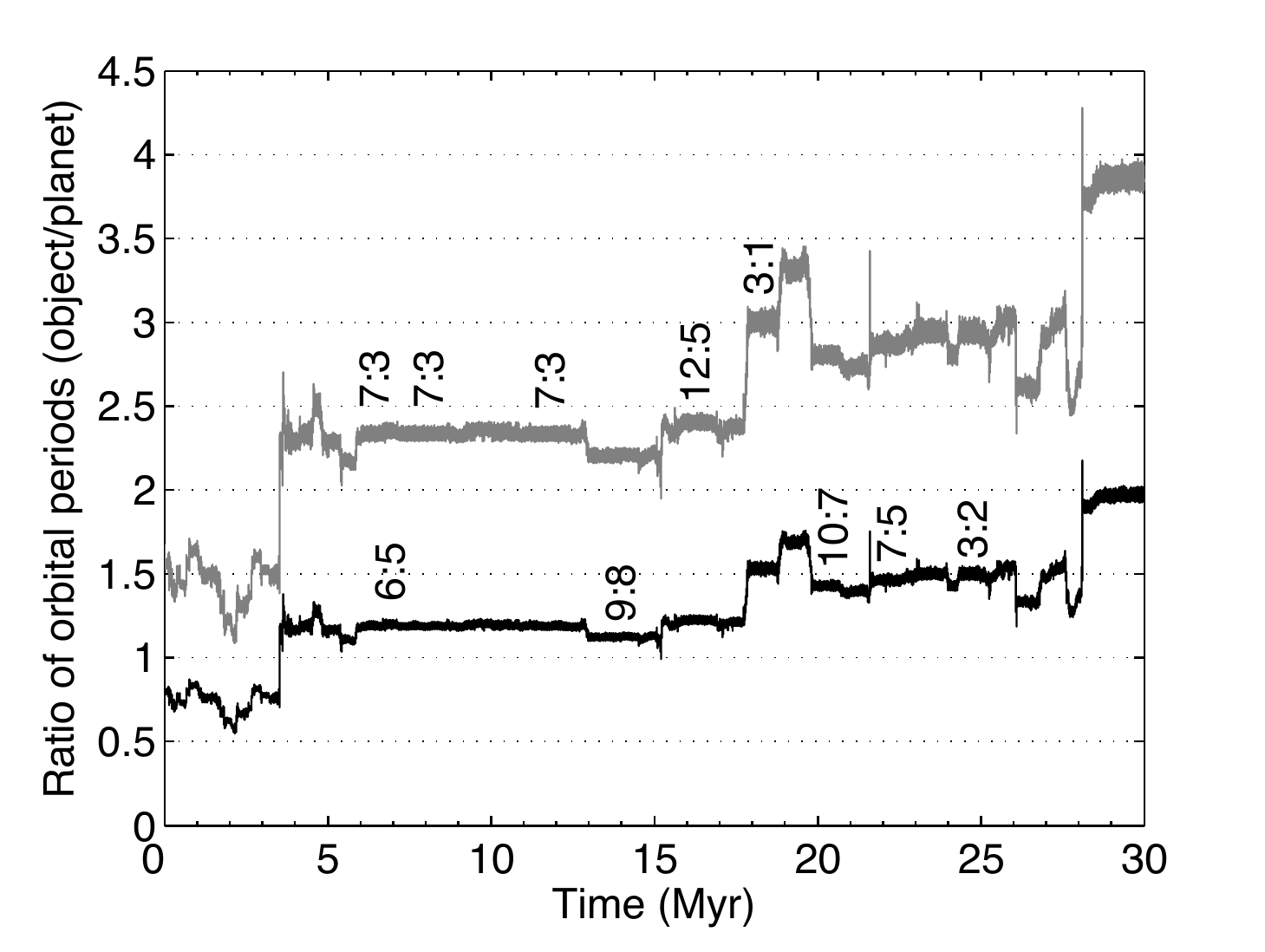}}
	\caption{A selection of mean motion resonances identified for 1998 TF35. The upper trace depicts resonances with Uranus; the lower trace corresponds to resonances with Neptune. }
	\label{fig:tf35_res}
\end{figure}

Many of the long-lived Centaurs spent considerable portions of time at constant semimajor axis.  In many cases, we have identified these time segments as mean motion resonances with the planets.  An example is presented in Fig.~\ref{fig:tf35_res}, on which we list the mean motion resonances identified for 1998 TF35 (also shown in the bottom panel of Fig.~\ref{fig:twosamples}).  We identified mean motion resonances by looking for small integer ratios $p:q$ between the orbital periods of the Centaur and the giant planets.  Each candidate pair $(p,q)$ was then used to define resonance angles of the form $\phi_a = p\lambda_{C} - q\lambda_{P} - (p - q)\varpi_{C}$ and $\phi_b = p\lambda_{C} - q\lambda_{P} - (p - q)\Omega_{C}$, where $\lambda$ is the mean longitude, $\varpi$ is the longitude of perihelion, $\Omega$ is the longitude of the ascending node, and the subscripts $C$ and $P$ refer to the Centaur and planet, respectively.  (It is possible to define many other combinations, involving the $\varpi$ and $\Omega$ angles for the planets; we did not consider those because such resonances are generally weaker due to the much smaller eccentricities and inclinations of the planets compared to those of the Centaurs.)  If either $\phi_a$ or $\phi_b$ librates, then $p:q$ is a mean motion resonance.
%

The close correspondence between resonance sticking/hopping and nonlinear plots of $\log\bar\sigma$ vs.~$\log{w}$ leads us to name the group of 10 objects with nonlinear plots as the resonance-hopping class. As noted above, this class does not have well-defined Hurst exponents. Most, though not all, of the long-lived Centaurs in our simulation exhibited resonance sticking during a significant fraction of their dynamical lifetimes.   Conversely, the short-lived Centaurs spent very little time in resonances.  From our results, resonance hopping is a relatively slow mechanism for chaotic orbital evolution, in contrast to the processes that cause the short-lived Centaurs to diffuse rapidly and either be ejected from the solar system or collide with a planet. Resonance hopping is further discussed in \citet{BM1997} and \citet{Getal2002}.  

\subsection{Exceptional Cases}
\label{exceptions}
\marginpar{\ \ \ \bf Fig.~8}

\begin{figure}[htb]
	\centering
		\includegraphics[width=1.00\textwidth]{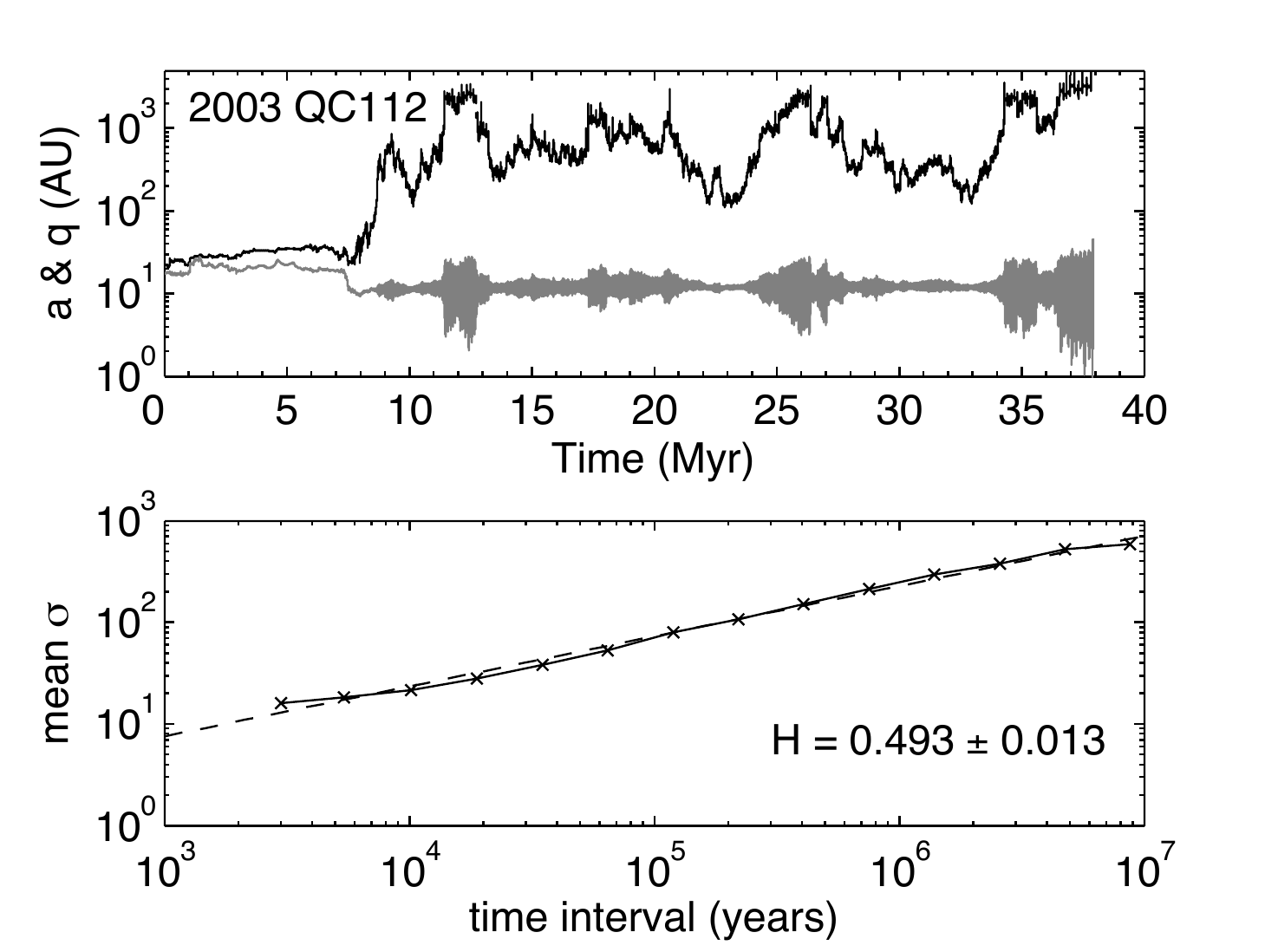}
	\caption{The upper panel shows semimajor axis and perihelion vs. time for 2003 QC112, a long-lived particle with exceptionally large standard deviation in semimajor axis; note the log scale on the vertical axis. The plot of log $\bar\sigma$ vs.~log $w$ is shown in the bottom panel.}
	\label{fig:03QC112}
\end{figure}

\begin{figure}[htb]
	\centering
		\includegraphics[width=1.00\textwidth]{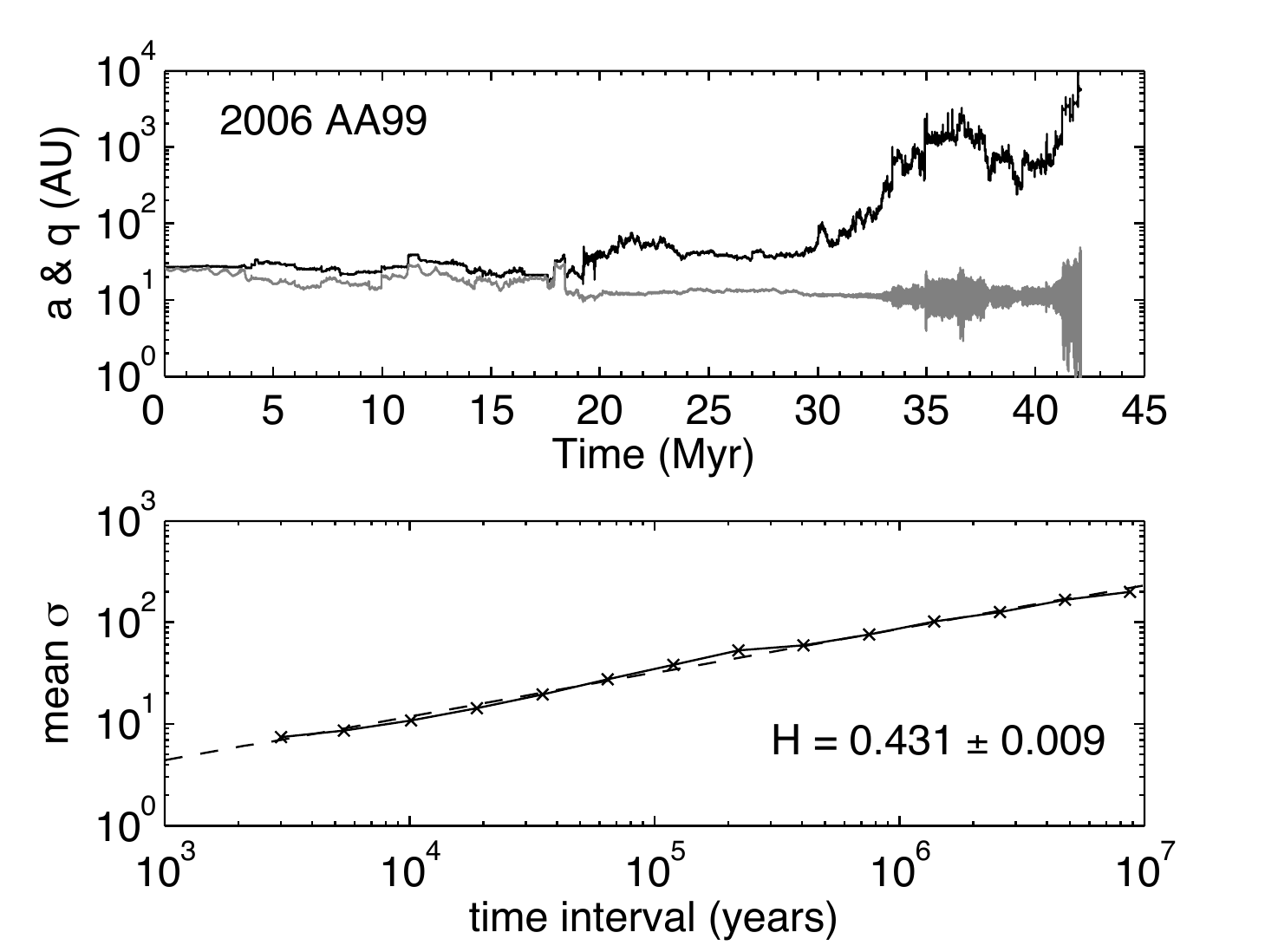}
	\caption{The upper panel shows semimajor axis and perihelion vs. time for 2006 AA99, a long-lived particle with exceptionally large standard deviation in semimajor axis; note the log scale on the vertical axis. The plot of log $\bar\sigma$ vs.~log $w$ is shown in the bottom panel.}
	\label{fig:2006AA99}
\end{figure}

\marginpar{\ \ \ \bf Fig.~9}
Two curves in the right-hand panel of Fig.~\ref{fig:slopes} lie well above the rest, with values of $\bar\sigma$ that are almost an order of magnitude larger than those of the other long-lived Centaurs. The individual results for these objects are shown in Fig.~\ref{fig:03QC112} and Fig.~\ref{fig:2006AA99}.
These objects, 2003 QC112 and 2006 AA99, spend much of their dynamical lifetimes at large semimajor axis, with large jumps near perihelion passage, and we have not detected any resonance sticking in their evolution. As seen in the bottom panels of each figure, their plots of $\log\bar\sigma$ vs.~$\log{w}$ are nearly linear.  These objects are members of the diffusion class rather than the resonance-hopping class, despite being long-lived. 

\marginpar{\ \ \ \bf Fig.~10}

\marginpar{\ \ \ \bf Fig.~11}

\begin{figure}[p]
	\centering
		\includegraphics[width=1.00\textwidth]{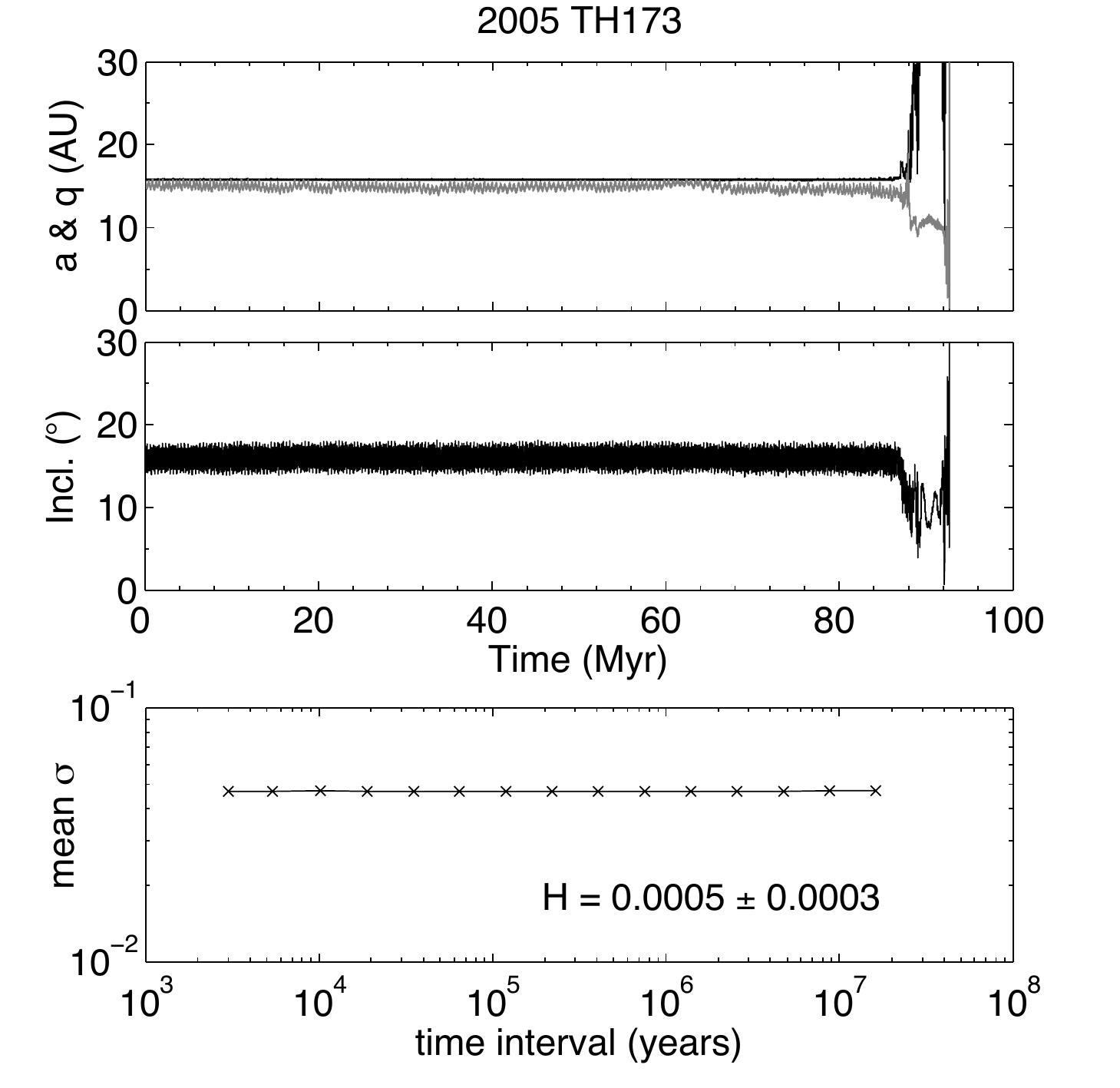}
	\caption{A quasistable object between Saturn and Uranus.  The top and middle panels show the time evolution of semimajor axis, perihelion, and inclination; the bottom panel shows the analysis of $\log\bar\sigma$ vs.~$\log{w}$ for the first 80 Myr of the particle's dynamical lifetime.}
	\label{fig:05TH173}
\end{figure}

\begin{figure}[p]
	\centering
		\includegraphics[width=1.00\textwidth]{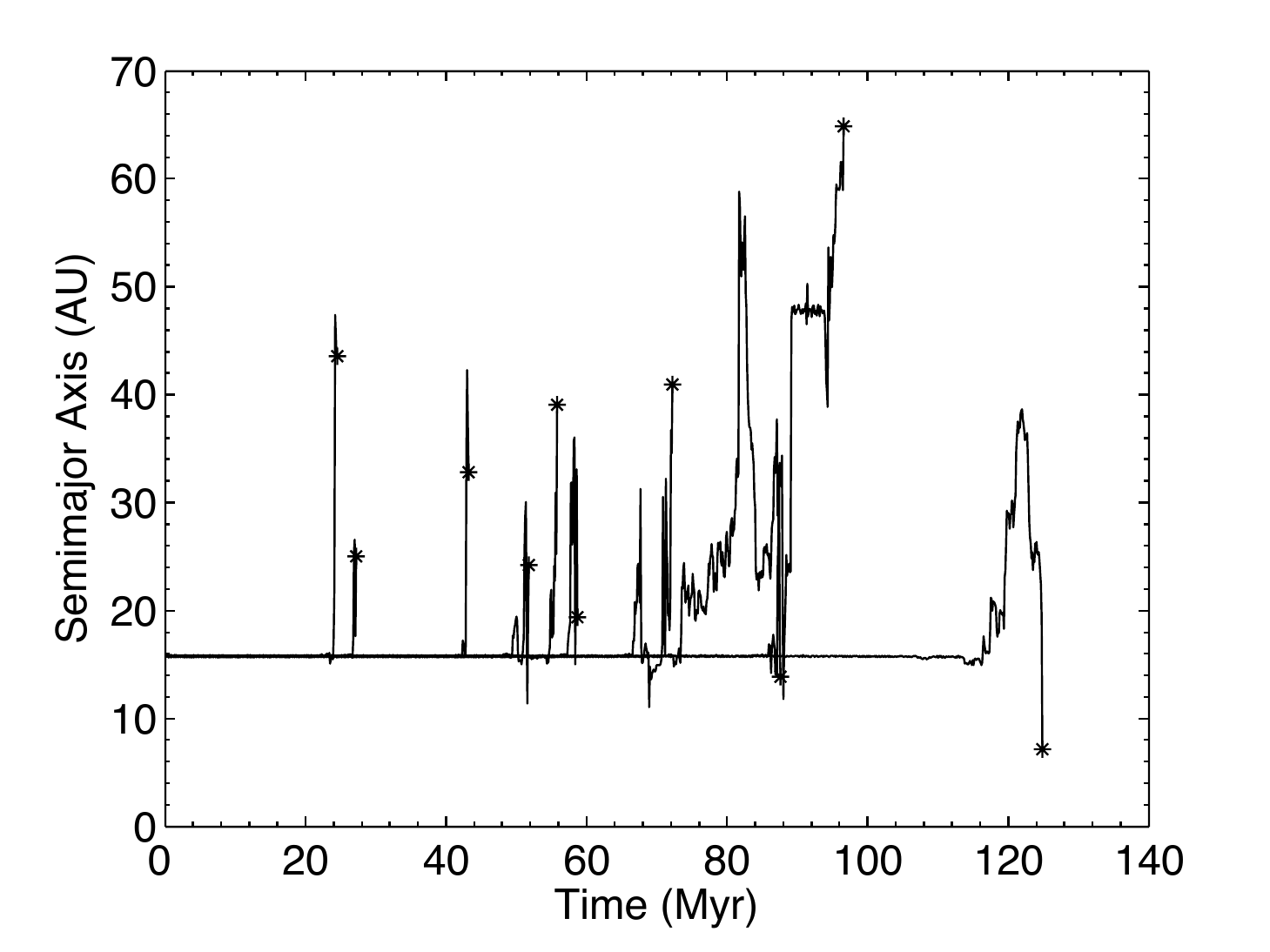}
	\caption{A plot of semimajor axis vs. time for an ensemble of ten clones of 2005 TH173. The asterisks indicate the endpoints of the orbital evolution for each clone.}
	\label{fig:TH173clones}
\end{figure}

Another object in our sample, 2005 TH173, remained at nearly constant semimajor axis for over 85 Myr (see Fig.~\ref{fig:05TH173}).  It follows a nearly circular, but inclined, orbit between Saturn and Uranus, with $a \simeq 15.8$ AU and inclination $i = 16^{\circ}$.  It is not in a mean motion resonance with either Saturn or Uranus.  Its $\log\bar\sigma$ vs.~$\log{w}$ curve for the first 80 Myr of its dynamical evolution is the lowermost line in the right-hand panel of Fig.~\ref{fig:slopes} and has slope zero.  This object could be a candidate for long sought but hitherto undiscovered long-lived orbits between the orbits of the giant planets \citep{H1997}.
To explore the stability of this orbit further, we numerically integrated an ensemble of 10 ``clones'' of 2005 TH173 for 125 Myr.  Each clone was randomly assigned an initial semimajor axis in the range $a_0 \pm 10^{-5}$ AU, where $a_0 = 15.724$ AU is the initial semimajor axis of the original object; all other initial orbital elements were identical to that of the original object.  For this integration, particles were removed from the simulation when they reached a heliocentric distance of 100 AU, and the evolved orbital elements were recorded every $10^5$ years.
As shown in Fig.~\ref{fig:TH173clones}, every clone survived at least 22 Myr in the same orbit, with the longest lasting more than 110 Myr.

\subsection{Correlations} \label{corr}
%
We have explored in some detail whether the initial conditions of the Centaurs are correlated with their dynamical behavior.   We measure the degree of correlation between two parameters by using Spearman's rank correlation coefficient, $r$: 
\begin{equation}
r = \frac{\left\langle  (x-\bar{x})(y-\bar{y}) \right\rangle}{\sqrt{\left\langle (x-\bar{x})^2 \right\rangle \left\langle(y-\bar{y})^2\right\rangle}},
\end{equation}
where $x$ and $y$ are the rank orders of the two parameters and $\bar x$ and $\bar y$ are their mean values.

First, we consider correlations amongst the initial conditions for each subset of our sample (the diffusing class and the resonance-hopping class), as these may provide clues regarding the possible source populations. 
The mean and standard deviation of the initial values of $a$, $e$, $i$, and $q$ for the two dynamical classes are given in Table \ref{table:means}. We see that the mean values of initial $a$, $i$, and $q$ are slightly larger for the resonance-hopping Centaurs than for the diffusion-dominated Centaurs, but the standard deviations within each group are large enough that these differences are not significant.  

\begin{table}[htbp] 
\caption{\small The mean (and standard deviation) of the initial orbital semimajor axis ($a$), eccentricity ($e$), inclination ($i$) and perihelion distance ($q$) For the two dynamical classes of Centaurs identified in our study. ``Number'' refers to the number of objects in each class.}\label{table:means} 
\begin{tabular}[t]{lccccc}

\hline
\small Class & Number & $a$ (AU) & $e$ & $i$ (deg) & $q$ (AU) \\
\hline
Diffusing & 51 & 19.3 $\pm$ 5.6 & 0.32 $\pm$ 0.18 & 13.3 $\pm$ 8.3 & 13.1 $\pm$ 5.2 \\
Resonance-hopping & 10 & 22.8 $\pm$ 3.6 & 0.29 $\pm$ 0.17 & 21.2 $\pm$ 12.2 & 15.8 $\pm$ 3.4 \\
	\hline	
	\end{tabular}
\end{table}

For the group of ten long-lived resonance-hopping Centaurs, we find weak positive correlations between $a$ and $e$ ($r = 0.52$) and between $e$ and $i$ ($r = 0.67$), and negative correlations between $e$ and $q$ ($r = -0.73$) and $i$ and $q$ ($r = -0.91$). 
 In contrast, we find no correlations among $a$, $e$ and $i$ for the diffusion-dominated Centaurs (maximum $|r| = 0.22$).  There is a weak positive correlation between $a$ and $q$ for this group, with $r = 0.67$, and a weak negative correlation between $e$ and $q$, with $r = -0.61$. 

We also find weak correlations between absolute magnitude and $e$, $i$, and $q$ for the resonance-hopping Centaurs, with $r = 0.72, 0.71$, and $-0.76$, respectively.  There are no such correlations for the diffusion-dominated group.  We see weak negative correlations between absolute magnitude and the Tisserand parameters with respect to Jupiter, Saturn, and Uranus, with $r \approx -0.7$ in each case, for the resonance-hopping group, but not for the diffusion-dominated group.  The strongest correlations are between $i$ and Tisserand parameter for the resonance-hopping Centaurs, with values of $-0.92$ for $r(i,T_{U})$ and $-0.95$ for $r(i,T_J)$ and $r(i,T_{S})$. For the diffusion-dominated Centaurs, the most significant correlation between $i$ and Tisserand parameter is only $r = -0.37$, for $r(i,T_U)$.  

These results suggest that the diffusing Centaurs are efficiently mixed (randomized) in orbital parameter space, but the resonance-hopping Centaurs are not so mixed.  

We also considered possible correlations with spectral colors of the Centaurs.  \citet{TBRP2008} report that the B--R colors of Centaurs are bimodal, with one gray and one red subpopulation, but that these colors show no correlation with orbital elements or absolute magnitude. Only four of our long-lived, resonance-hopping class have published colors; this is too small a data set to test for correlations between color and dynamical class from our sample. 

The Hurst exponent provides a measure of the rate of transport for a particle, with large values indicating rapid transport and small values indicating slow transport.  \citet{DB2007} found that mean lifetimes for subsets of their sample of Centaurs depended on initial inclination and perihelion distance.  We could thus expect that the Hurst exponent would be correlated with $i$ and $q$.  However, we found no significant correlations between the Hurst exponent and any orbital elements or Tisserand parameters within the diffusing group.  
This result may simply reflect the fact that properties of mean values of an ensemble need not apply to time series of individual particles. 

\subsection{Link to Jupiter-family Comets}
\label{jfcs}
We have also investigated the dynamical link between Centaurs and the Jupiter-family comets (JFCs).  Because many of the particles in our simulation reach high inclinations, the definition of JFCs as given by \citet{LD1997}, $2<T_J<3$, is not sufficient to identify objects whose dynamics are dominated by Jupiter.  We therefore adopted a modified definition, as proposed by \citet{GMV2008}, which includes a condition on perihelion distance: $q < 7.35$ AU.  This distance is midway between the orbits of Jupiter and Saturn. 

\marginpar{\ \ \ \bf Fig.~12}

\begin{figure}[htbp]
	\centering
		\includegraphics[width=1.00\textwidth]{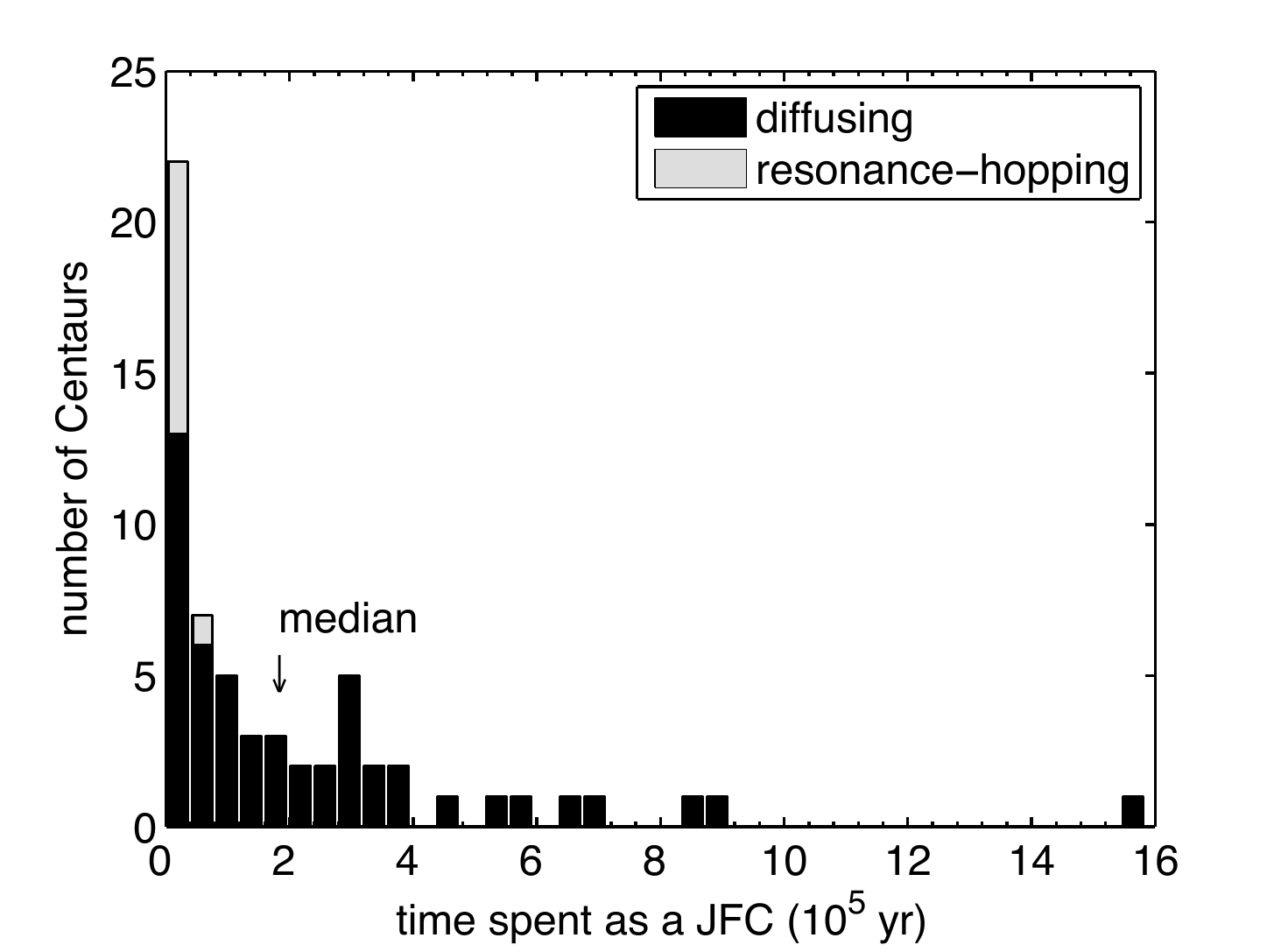}
	\caption{Histogram of the time spent as a JFC for the particles in our simulation.  Members of the diffusing class are shown in black, while resonance-hopping particles are shown in gray.}
	\label{fig:jfctimes}
\end{figure}

Using the modified definition of JFCs, we found that 47 of the 62 particles (76\%) spent part of their dynamical lifetimes as JFCs.  A histogram of the time spent as a JFC for our sample of particles is shown in Fig.~\ref{fig:jfctimes}. The median time spent as a JFC was $1.6 \times 10^5$ years, comparable to the dynamical lifetimes of JFCs found in other work \citep{LD1994}.  All but one of these 47 were members of the diffusing class of Centaurs.  In contrast, 9 out of 10 of the resonance-hopping class never became JFCs.  This suggests that the JFCs are supplied by a subset of the Centaurs, the diffusing dynamical class.  We note that, in our sample, the initial perihelion distance of the objects which do become JFCs ranges from 5.723 AU to 22.827 AU, a span that also encompasses the range of initial perihelion distance for the resonance-hopping Centaurs.  Initial $q$ is thus not a predictor of whether or not a Centaur will become a JFC during its dynamical lifetime.  Furthermore, the values of the Hurst exponents for the objects which become JFCs range from 0.21 to 0.95, with a roughly normal distribution.  This means that both fast and slow diffusers become JFCs.

\section{Summary and Conclusions}
\label{conc}
Our analysis of the long term orbital evolution of known Centaurs shows that these objects can be classified into two dynamical classes: one is characterized by diffusive evolution of semimajor axis and the other is dominated by resonance hopping.  The two classes can be objectively and quantitatively distinguished by calculating Hurst exponents from the time series of their semimajor axes.  Objects in the diffusing class have well-defined Hurst exponents, while objects in the resonance-hopping class do not.  This dynamical classification is strongly correlated with dynamical lifetime: all ten of the Centaurs in the resonance-hopping category survived at least 40 Myr, and eight survived the full 100 Myr; in contrast, 46 of the 51 diffusing Centaurs were ejected or collided with a massive body within 22 Myr, more than half of these within 6 Myr. 

The resonance-hopping class of Centaurs exhibits weak correlations among the initial orbital elements $a$, $e$, $i$, and $q$, as well as absolute magnitude; no such correlations are found in the diffusive class of Centaurs. Within the diffusive class, we found no significant correlations between the Hurst exponent and any orbital parameters. These results suggest that the diffusing Centaurs are efficiently mixed (randomized) in orbital parameter space, but the resonance-hopping Centaurs are less so; the latter may be preserving some memory of their source.   

The diffusive class has mean values of $a$, $q$ and $i$ slightly smaller than that of the resonance-hopping class, but the differences are less than the standard deviations, hence not significant.  (We note that all of the resonance-hopping Centaurs initially lie on orbits exterior to Saturn, with semimajor axes between 18 and 30 AU and perihelion in the range 11.9--18.9 AU; in some contrast, the diffusing group has initial semimajor axes in the range 7.9 AU to 30 AU and perihelion in the range 5.6--25.6 AU. Future studies with larger samples could determine if there is significant systematic difference between the two groups' initial orbits.) Our simulations indicate that the diffusing class of Centaurs are far more likely to evolve into Jupiter family comet-type orbits than the resonance-hopping class of Centaurs.  There are currently insufficient data on the colors of Centaurs to determine whether the two dynamical classes exhibit different color trends.   More work with larger samples needs to be done to evaluate the significance of these correlations (or lack thereof), and to understand the origins of the diffusive and resonance-hopping dynamical classes.   

\subsection*{Acknowledgments}
This research was supported in part by NASA's Outer Planets Research Program,
grant nos.~NNG05GH44G and NNX08AQ65G.  We thank reviewers A.~Brunini and L.~Dones for comments that helped to improve this paper.
  

\bigskip

\bibliographystyle{elsarticle-harv}
\bibliography{kborefs}

\clearpage

\clearpage
\clearpage
\clearpage
\clearpage
\clearpage
\clearpage
\clearpage
\clearpage
\clearpage
\clearpage
\clearpage

\clearpage

\begin{longtable}{lcccccrc} 
\caption{\small Data and classification of the Centaurs in our study. The data from the MPC (accessed on 06 March 2007) are the values of orbital semimajor axis ($a$), eccentricity ($e$), inclination ($i$) and perihelion distance ($q$), the absolute visual magnitude, and the length of observational arc ("`Opps."' is the number of observed oppositions, if more than 1; values in parentheses are the observational arcs (in days) for single oppositions).  The last column, ``Class", gives the dynamical class determined in our analysis (section~\ref{anal}): D refers to the diffusing class, R refers to the resonance-hopping class, Q refers to a quasi-stable orbit, and NT denotes a Neptune Trojan.}
\label{table:ics} \\

\hline
\small Name & $a$ (AU) & $e$ & $i$ (deg) & $q$ (AU) & Mag. & Opps. & Class  \\
\hline
\endfirsthead

\multicolumn{8}{c}%
{{\bfseries \tablename\ \thetable{} -- continued from previous page}} \\
\hline
Name & $a$ (AU) & $e$ & $i$ (deg) & $q$ (AU) & Mag. & Opps. & Class  \\
\hline 
\endhead

\hline \multicolumn{8}{r}{{Continued on next page}} \\ \hline
\endfoot

\hline \hline
\endlastfoot

1977 UB & 13.701 & 0.381 & 6.9 & 8.479 & 6.5 & 29 & D \\
1992 AD & 20.411 & 0.572 & 24.7 & 8.743 & 7 & 14 & D \\
1993 HA2 & 24.662 & 0.52 & 15.6 & 11.832 & 9.6 & 7 & D \\
1994 TA & 16.722 & 0.304 & 5.4 & 11.645 & 11.5 & 5 & D \\
1995 DW2 & 25.196 & 0.25 & 4.1 & 18.905 & 8 & 6 & R \\
1995 GO & 18.015 & 0.621 & 17.6 & 6.834 & 9 & 11 & D \\
1995 SN55 & 23.564 & 0.663 & 5 & 7.938 & 6 & (36d) & D \\
1996 AR20 & 15.197 & 0.627 & 6.2 & 5.666 & 14 & (13d) & D \\
1996 RX33 & 23.868 & 0.204 & 9.4 & 19.009 & 9.3 & (13d) & D \\
1997 CU26 & 15.854 & 0.175 & 23.4 & 13.082 & 6.4 & 9 & D \\
1998 QM107 & 19.997 & 0.137 & 9.4 & 17.25 & 10.4 & 5 & R \\
1998 SG35 & 8.382 & 0.308 & 15.6 & 5.8 & 11.3 & 8 & D \\
1998 TF35 & 26.082 & 0.378 & 12.7 & 16.22 & 9.3 & 4 & R \\
1999 HD12 & 21.322 & 0.583 & 10.1 & 8.898 & 12.8 & (50d) & D \\
1999 JV127 & 16.724 & 0.359 & 25.5 & 10.719 & 10.4 & (8d) & D \\
1999 UG5 & 11.769 & 0.383 & 5.3 & 7.259 & 10.1 & 7 & D \\
1999 XX143 & 17.934 & 0.464 & 6.8 & 9.621 & 8.6 & 4 & D \\
2000 CO104 & 24.173 & 0.147 & 3.1 & 20.611 & 10.1 & 3 & D \\
2000 EC98 & 10.773 & 0.456 & 4.3 & 5.856 & 9.5 & 7 & D \\
2000 FZ53 & 23.869 & 0.479 & 34.8 & 12.438 & 11.4 & 3 & R \\
2000 GM137 & 7.904 & 0.122 & 15.8 & 6.943 & 14.3 & 3 & D \\
2000 QC243 & 16.483 & 0.2 & 20.7 & 13.187 & 7.6 & 7 & D \\
2000 SN331 & 17.988 & 0.201 & 14.7 & 14.374 & 10.9 & (1d) & R \\
2001 BL41 & 9.759 & 0.293 & 12.5 & 6.899 & 11.7 & 4 & D \\
2001 KF77 & 26.161 & 0.243 & 4.4 & 19.812 & 9.5 & 4 & D \\
2001 PT13 & 10.622 & 0.199 & 20.4 & 8.51 & 9 & 7 & D \\
2001 SQ73 & 17.411 & 0.176 & 17.5 & 14.348 & 9.6 & 4 & D \\
2001 XA255 & 29.967 & 0.688 & 12.7 & 9.349 & 11.2 & 5 & D \\
2001 XZ255 & 15.933 & 0.034 & 2.6 & 15.393 & 11.1 & 2 & D \\
2002 CA249 & 20.713 & 0.385 & 6.4 & 12.744 & 12 & (37d) & D \\
2002 CB249 & 28.421 & 0.511 & 14 & 13.899 & 9.8 & (35d) & D \\
2002 DH5 & 22.116 & 0.367 & 22.5 & 13.997 & 10.2 & 3 & D \\
2002 FY36 & 28.969 & 0.114 & 5.4 & 25.654 & 8.4 & (51d) & D \\
2002 GB10 & 25.246 & 0.398 & 13.3 & 15.2 & 7.8 & 8 & D \\
2002 GO9 & 19.537 & 0.281 & 12.8 & 14.055 & 9.1 & 6 & D \\
2002 GZ32 & 23.196 & 0.223 & 15 & 18.033 & 6.8 & 5 & D \\
2002 KY14 & 12.724 & 0.137 & 17 & 10.98 & 9.9 & (26d) & D \\
2002 PQ152 & 25.7 & 0.199 & 9.4 & 20.584 & 8.6 & (84d) & D \\
2002 QX47 & 25.491 & 0.375 & 7.3 & 15.933 & 8.9 & 2 & D \\
2002 TK301 & 16.1 & 0.132 & 24.4 & 13.969 & 13.4 & (1d) & D \\
2002 VR130 & 23.833 & 0.382 & 3.5 & 14.725 & 11 & 4 & D \\
2002 VG131 & 17.487 & 0.15 & 21.7 & 14.869 & 11.2 & (23d) & D \\
2003 CO1 & 20.932 & 0.478 & 19.7 & 10.926 & 8.9 & 4 & D \\
2003 KQ20 & 10.563 & 0.194 & 5.7 & 8.515 & 13.1 & (1d) & D \\
2003 LH7 & 16.964 & 0.292 & 21.2 & 12.002 & 12.5 & (1d) & D \\
2003 QC112 & 22.083 & 0.21 & 16.7 & 17.449 & 8.7 & (60d) & D \\
2003 QD112 & 19.013 & 0.582 & 14.5 & 7.939 & 10.9 & 2 & D \\
2003 QN112 & 25.115 & 0.333 & 7.9 & 16.743 & 12.9 & (54d) & D \\
2003 QP112 & 21.129 & 0.329 & 31.2 & 14.184 & 12.7 & (54d) & R \\
2003 UW292 & 18.109 & 0.131 & 21 & 15.728 & 8.4 & (28d) & R \\
2003 UY292 & 21.864 & 0.272 & 8.6 & 15.913 & 10.2 & (29d) & D \\
2003 WL7 & 20.077 & 0.256 & 11.2 & 14.947 & 8.7 & 4 & D \\
2004 CJ39 & 12.959 & 0.482 & 3.6 & 6.715 & 14 & (52d) & D \\
2004 QQ26 & 23.068 & 0.153 & 21.4 & 19.535 & 9.4 & 2 & D \\
2004 XQ190 & 23.058 & 0.01 & 6.3 & 22.824 & 12 & (2d) & D \\
2005 RL43 & 24.466 & 0.042 & 12.3 & 23.429 & 8.4 & 3 & R \\
2005 RO43 & 28.772 & 0.52 & 35.5 & 13.822 & 7.3 & 3 & R \\
2005 TH173 & 15.724 & 0.014 & 15.7 & 15.5 & 11 & (17d) & Q \\
2005 UJ438 & 17.525 & 0.529 & 3.8 & 8.252 & 10.5 & 5 & D \\
2005 VB123 & 17.744 & 0.009 & 38.9 & 17.592 & 10.2 & (29d) & D \\
2006 AA99 & 26.859 & 0.045 & 33.4 & 25.639 & 11.9 & (1d) & D \\
2006 RJ103 & 29.973 & 0.028 & 8.2 & 29.12 & 7.5 & 2 & NT \\ 
2006 SX368 & 22.134 & 0.459 & 36.3 & 11.969 & 9.5 & (74d) & R \\
\hline
\end{longtable}
\end{document}